\documentclass{ecai}  

\usepackage{graphicx}
\usepackage{latexsym}
\usepackage{amsmath}
\usepackage{amsthm}
\usepackage{booktabs}
\usepackage{algorithm}
\usepackage{algpseudocode}
\usepackage{amsfonts}
\usepackage{enumitem}
\usepackage{bm}
\usepackage{multirow}
\usepackage{enumitem}
\usepackage{bm}
\usepackage{pifont}
\usepackage[switch]{lineno}

\usepackage[justification=justified]{caption}

\usepackage{threeparttable}

\DeclareMathOperator*{\argmax}{arg\,max}
\DeclareMathOperator*{\argmin}{arg\,min}

\newcommand{\abbr}{\text{XGBD}}

\title{\(\abbr\): Explanation-Guided Graph Backdoor Detection}
\begin{document}

\begin{frontmatter}

\author[1]{\fnms{Zihan}~\snm{Guan}}
\author[2]{\fnms{Mengnan}~\snm{Du}}
\author[1]{\fnms{Ninghao}~\snm{Liu} \thanks{Corresponding Author. Email: ninghao.liu@uga.edu.}}
\address[1]{School of Computing, University of Georgia, USA}
\address[2]{Department of Data Science, New Jersey Institute of Technology, USA}

\begin{abstract}
Backdoor attacks pose a significant security risk to graph learning models. Backdoors can be embedded into the target model by inserting backdoor triggers into the training dataset, causing the model to make incorrect predictions when the trigger is present. To counter backdoor attacks, backdoor detection has been proposed. An emerging detection strategy in the vision and NLP domains is based on an intriguing phenomenon: when training models on a mixture of backdoor and clean samples, the loss on backdoor samples drops significantly faster than on clean samples, allowing backdoor samples to be easily detected by selecting samples with the lowest loss values. However, the ignorance of topological feature information on graph data limits its detection effectiveness when applied directly to the graph domain. To this end, we propose an explanation-guided backdoor detection method to take advantage of the topological information. Specifically, we train a helper model on the graph dataset, feed graph samples into the model, and then adopt explanation methods to attribute model prediction to an important subgraph. We observe that backdoor samples have distinct attribution distribution than clean samples, so the explanatory subgraph could serve as more discriminative features for detecting backdoor samples. Comprehensive experiments on multiple popular datasets and attack methods demonstrate the effectiveness and explainability of our method. Our code is available: \url{https://github.com/GuanZihan/GNN_backdoor_detection}.
\end{abstract}

\end{frontmatter}

\maketitle

\section{Introduction}
Graph machine learning has recently gained popularity due to its high expressibility in a wide range of graph-based applications such as social networks~\cite{Wu_Lian_Xu_Wu_Chen_2020}, physical systems~\cite{Sanchez2018}, and biomedical analysis~\cite{NIPS2017_f5077839}. Graph neural networks (GNNs) achieve state-of-the-art performance in various graph learning tasks~\cite{wu2020comprehensive,zhou2020graph,zhang2020deep,shi2023engage}. In practice, due to the complexity of graph data and the difficulty of data collection, it is common for users to directly use graph datasets published by a third party~\cite{tudataset} or from untrusted Internet sources, which could lead to potential security problems. For example, a malicious entity could launch backdoor attacks by injecting predefined triggers (e.g., small subgraphs) into victim samples and changing their labels to target values. If users train their models on the poisoned dataset, a mapping from the trigger patterns to the target labels will be learned. 
Then, if the model is deployed to real-world applications, the malicious entity could manipulate model behaviors by feeding poisoned inference samples into the model.

There has been a plethora of prior work on backdoor detection in the vision and NLP domains~\cite{activation_cluster,strip,neuron_inspect,dong2021,AEVA2020,2022Xiang,hayase2022few,chen2023cleanimage,yu2023backdoor,chou2022backdoor}. However, backdoor detection on graph data has not been well explored. Although some work~\cite{activation_cluster,abl} originally proposed in the vision domain can be applied directly to the graph domain, their effectiveness is limited as they ignore the topological information of the graph data. To the best of our knowledge, only a few preliminary works focus on backdoor detection on graphs~\cite{gnnDefense2022}. However, it requires an additional validation set to determine the detection boundary, which is not always realistic in practice.

A recently observed phenomenon provides opportunities for detecting backdoor samples: When training models on a mixture of backdoor samples and clean samples, the loss on backdoor samples tends to decrease significantly fast in the early epochs, while the loss on clean samples tends to decrease steadily. Therefore, the backdoor samples could be easily identified by selecting samples with the top-k lowest loss values. The method has been shown to be effective in the vision domain~\cite{abl}. However, the method does not perform effectively in the graph domain due to the following challenges: 1) How to take advantage of the topological information in the graph data? 2) How to enlarge the loss gap between backdoor samples and clean samples in the graph domain? 

To this end, we develop an eXplanation-Guided Backdoor Detection (\(\abbr\)) method. It is motivated by the following empirical findings: Given backdoor samples, neural networks are prone to solely rely on simple trigger patterns to make predictions, whereas given clean samples, they tend to rely on more complex feature patterns. 
It is easier for an explanation method to capture the trigger patterns of backdoor samples than to find all the key features of clean samples, thus creating a gap between the two types of samples. 
Therefore, given a training dataset which could contain backdoor samples, we train a graph neural network on the dataset, and apply explanation methods on each of the training samples to extract an explanatory subgraph. 
As a result, the backdoor samples could be detected based on the model's loss value with the explanatory subgraph as input. If the loss value is less than some threshold $\tau$, i.e., the explanatory subgraph well preserves the core information of the overall graph, the input sample is regarded as containing backdoors.
\begin{figure}[!t]
    \centering
    \includegraphics[width=0.49\textwidth]{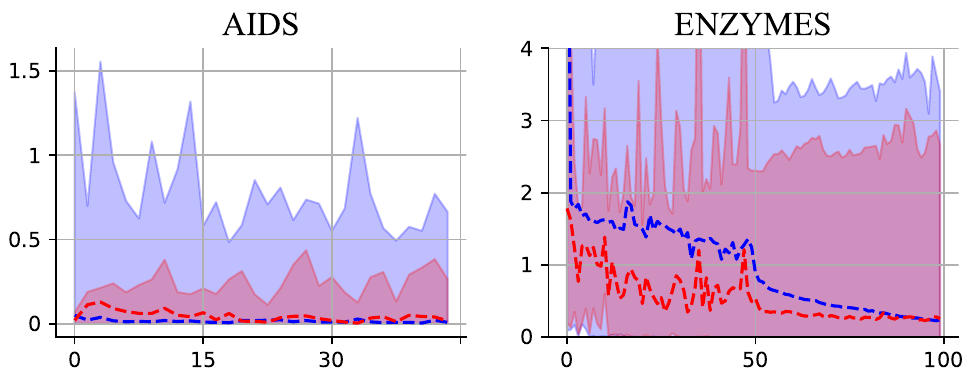}
    \vspace{-10pt}
    \caption{The average training loss curve for clean samples and backdoor samples on different datasets. The blue and red bands denote the range of loss values for clean samples and backdoor samples, respectively. The $x$-axis denotes the training epochs, and the $y$-axis denotes the training loss value. The gap between the loss curves of clean and backdoor samples is \textit{not} significant enough to effectively distinguish them for backdoor samples detection.}
    \label{fig:pre_experiment}
\end{figure}

In summary, our contributions lie in the following aspects:
\begin{itemize}
    \item We propose a backdoor detection method for graph data. Compared to previous work~\cite{gnnDefense2022}, our method does not require an additional validation dataset.
    \item We show that explanation methods are effective feature filters in distinguishing between clean samples and backdoor samples when integrated with our carefully crafted constraints and detectors (Section 3.2).
    \item We conduct comprehensive experiments with rich ablation studies, where our proposed method achieves satisfactory performance across various datasets (Section 4).
\end{itemize}

\section{Preliminaries}
\subsection{Graph Neural Networks}
\noindent \textbf{Notations.} Let $\mathcal{G}=(\mathcal{V}, \mathcal{E}, \mathcal{X})$ denote a graph with a finite node set $\mathcal{V}$ of size $n$, an edge set $\mathcal{E}$, and a node feature matrix $\mathcal{X} \in \mathbb{R}^{n \times m}$. $A \in \mathbb{R}^{n\times n}$ is the corresponding adjacent matrix, where $A_{i,j}\in \{0,1\}, \forall i, j \in \mathcal{V}$, indicates whether there is an edge between node $i$ and node $j$. The neighbors of a node $v \in \mathcal{V}$ are denoted as $\mathcal{N}(v) = \{u \in \mathcal{V}|(u, v) \in \mathcal{E}\}$; the neighbors of a subset $S \subset \mathcal{V}$ are defined as $\mathcal{N}(S) = \bigcup_{v \in S} N(v)$. For simplicity, we also denote $\mathcal{X}(S) \in \mathbb{R}^{|S|\times m}$ as the feature matrix of the subset $S$.

\vspace{4pt}
\noindent \textbf{Graph Neural Networks (GNNs).} GNNs have become the state-of-the-art approach for modeling graph data. Typically, GNNs take a graph $\mathcal{G}$ as input and use the message-passing mechanism to learn its embedding and prediction. Without loss of generality, the message passing towards node $v$ at layer $l$ is defined as below:
\begin{equation}
     h_{v}^{l+1} = \sigma(\bm{a}_v^{l} W^l),\,\,
    \bm{a}_v^{l} = g^{l}(h_v^{l}, \{h_u^{l}:u\in \mathcal{N}(v)\}),
\end{equation}
where $h_{v}^{l}$ is the intermediate embedding of node $v$ at layer $l$, $W^l$ denotes the trainable parameter at layer $l$, $\sigma$ denotes the activation function, and $g^l$ is the aggregation function at layer $l$ that collects information from neighbors $\mathcal{N}(v)$.

The node embedding in the final layer could be used for downstream tasks such as node classification. For graph-level tasks such as graph classification, a read-out function is further added to aggregate all node-level embeddings to obtain a graph-level embedding $H_{\mathcal{G}}$ for the graph $\mathcal{G}$:
\begin{equation}
    H_{\mathcal{G}} = \text{READOUT}(\{h_{v}^L:v\in \mathcal{V}\}),
\end{equation}
where the $\text{READOUT}(\cdot)$ function can be a simple average function or a carefully crafted pooling function~\cite{gao2019graph}.

\vspace{4pt}
\noindent \textbf{Backdoor Attacks.}
The goal is to embed backdoors into the target model, where the model is expected to perform normally on clean input, but behave incorrectly if a trigger is present in the input graph during the inference stage. The backdoor triggers could either be injected into the target label by poisoning the dataset and luring the users to train on the dataset, or by directly modifying the model weights. In this paper, we consider the former situation.

\vspace{4pt}
\noindent \textbf{Graph Classifier Training in Backdoor Attacks.} A GNN-based classifier $f_{\theta}: \mathcal{G} \rightarrow \mathbb{R}^C$ maps the input graph $\mathcal{G}$ into its prediction distribution, where $C$ is the number of classes. Let $\mathcal{D} = \{(\mathcal{G}_1, y_1), (\mathcal{G}_2, y_2),...,(\mathcal{G}_n, y_n)\}$ denote the dataset, where $\mathcal{G}_i$ and $y_i$ denote the $i$-th graph and its ground-truth label. In a standard machine learning paradigm, $\mathcal{D}$ is usually divided into two parts: $\mathcal{D}_{train}$ and $\mathcal{D}_{test}$, for training and testing, respectively. In the context of backdoor attacks, $\mathcal{D}_{train}$ are injected with backdoor samples, where $\mathcal{D}_{train} = \mathcal{D}_{train}^{poi} \cup \mathcal{D}_{train}^{clean}$, with $\mathcal{D}_{train}^{poi}$ denoting the backdoor subset and $\mathcal{D}_{train}^{clean}$ denoting the clean subset. In this paper, $\mathcal{D}_{train}$ specifically denotes a poisoned dataset with an injection ratio of $\eta = |\mathcal{D}_{train}^{poi}|/|\mathcal{D}_{train}|$. Then a target GNN classifier $f_{\theta}$ is trained over $\mathcal{D}_{train}$ as follows,
\begin{equation}
    \theta = \argmin_{\theta'} \mathbb{E}_{(\mathcal{G}, y) \sim \mathcal{D}_{train}} \ell (f_{\theta'}, \mathcal{G}, y),
\end{equation}
where $\ell(\cdot)$ is the loss function on a single graph sample.

\subsection{Threat Model}
We define the threat model in two aspects: attacker and defender, following previous work on backdoor detection~\cite{activation_cluster,spectre2021}.
\begin{figure*}[!t]
    \centering
    \includegraphics[width=0.94\textwidth]{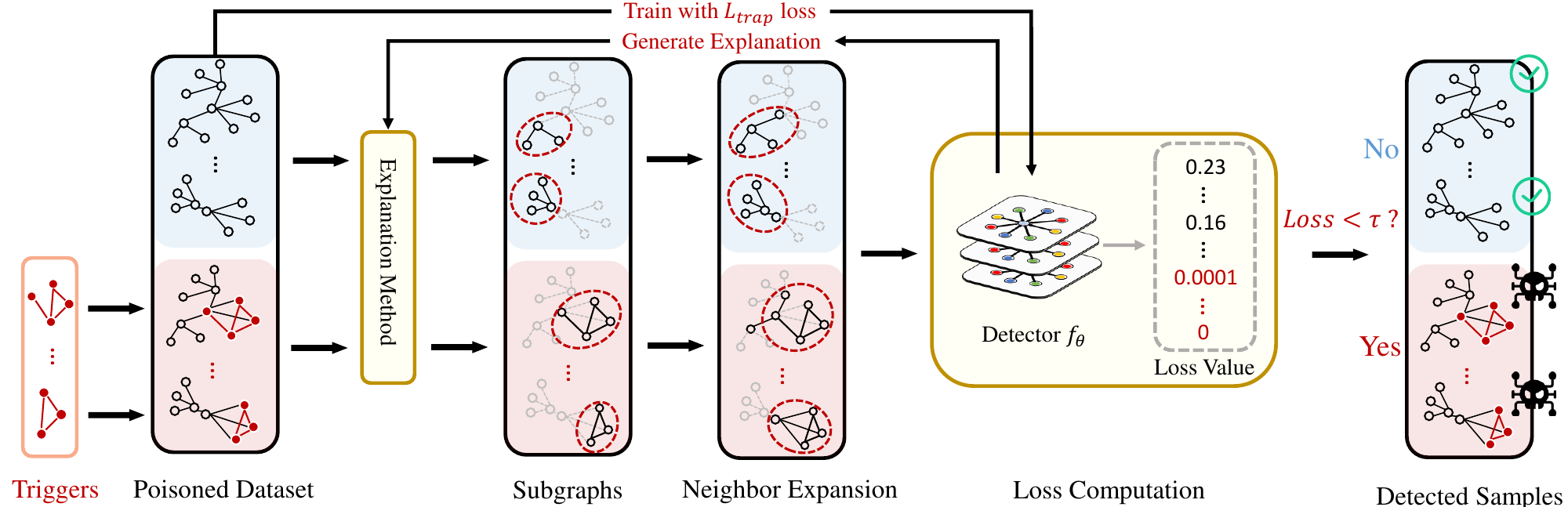}
    \caption{The pipeline of our proposed XGBD framework. After training $f_{\theta}$ with $L_{trap}$, each graph in the dataset goes through the explanation method, generating a set of explanatory subgraphs. After neighbor expansion, the subgraphs are fed back into $f_{\theta}$, where samples with loss value below threshold $\tau$ are detected as backdoor samples. }
    \label{fig:pipeline}
\end{figure*}

\vspace{4pt}
\noindent \textbf{Attacker.}
We assume that the attackers have white-box access to the training dataset $\mathcal{D}_{train}$, i.e., they can inject backdoor samples into the training dataset. The attackers can construct backdoor samples with four parameters: trigger size, trigger density, trigger synthesis method, and injection ratio, where trigger size characterizes the ratio of nodes to be poisoned, the trigger density describes the probability of choosing each possible edge for the subgraph trigger, trigger synthesis method describes how the trigger is generated, and the injection ratio is the parameter $\eta$ described above. To poison the training dataset, a random subgraph $\mathcal{S} \subset \mathcal{G}$ is chosen for each victim sample ($\mathcal{G}, y$) and replaced by the subgraph trigger $t$, along with its label $y$ altered to the target label $\hat{y}$. We assume that the attackers only poison the dataset but do not train the models on their own.

\vspace{4pt}
\noindent \textbf{Defender.}
We assume that the defender has full access to the training dataset $\mathcal{D}_{train}$. Given the $\mathcal{D}_{train}$, the defender's goal is to detect the backdoor samples. The defenders are assumed to have the flexibility to pursue different methodologies in achieving this goal. For example, they could train their own models or detectors over $\mathcal{D}_{train}$.

\section{Methodology}
\subsection{A Na\"ive Detection Method}
When training the neural network on a mixture of backdoor samples and clean samples, it was recently discovered that the losses on backdoor samples tend to drop significantly faster than clean samples in the early epochs~\cite{abl}. Based on the phenomenon, a na\"ive method of backdoor detection is to select the top-k samples from $D_{train}$ with the lowest training loss during the early training epochs. This intuitive method has been shown to be effective in image classification~\cite{abl}. However, its effectiveness is diminished when applied directly to graph learning tasks. Figure~\ref{fig:pre_experiment} presents a preliminary example showing the average training loss curve for the backdoor and clean samples for different epochs. In the experiment, a GIN model~\cite{gin2018xu} is trained on a poisoned AIDS/ENZYMES training dataset $D_{train}$ with injection ratio $\eta = 0.1$. We track the average training loss for the backdoor samples and clean samples shown in Figure~\ref{fig:pre_experiment}, where the blue line represents the clean sample curve and the red line represents the backdoor sample curve. However, the gap between the two curves is not significant, so the backdoor samples cannot be easily isolated. If we try to isolate 10\% samples with the lowest values after training with one epoch (since the average loss gap is the largest after the first epoch), the precision of the isolated backdoor samples is only $28.57\%/38.23\%$ for the two datasets, respectively. 
The possible reason could be that subgraph triggers contain topological information, making them harder to be learned by the model, compared to triggers in the image domain (e.g., mosaic patches). To solve the problem, the question is \emph{\textbf{how to enlarge the loss difference between backdoor samples and clean samples.}}

\subsection{Topological Feature Filtering with Explanation}
To this end, we propose to utilize explainability to guide the detection of backdoor samples. Specifically, we adopt post-hoc local explanation methods as feature filters to distinguish backdoor samples from clean samples. Given a prediction $f_{\theta}(\mathcal{G})$, where $f_{\theta}$ is a trained GNN model and $\mathcal{G}$ is an input graph, the goal of the explanation is to extract a subgraph $\mathcal{G}^* \subset \mathcal{G}$ that preserves the most important information of $\mathcal{G}$ to the prediction. In general, we try to solve the following optimization problem via graph model explanation:
\begin{equation}
    \mathcal{G}^* = \argmax_{|\mathcal{G}'| \leq \Omega}\, \text{Score}(\mathcal{G}, \mathcal{G}', f_{\theta}) ,
\end{equation}
where $\Omega$ is the upper bound on the size of the explanatory subgraph $\mathcal{G}^*$, and $\text{Score}(\cdot)$ is a scoring function that measures the importance of the subgraph $\mathcal{G}'$ to the original graph $\mathcal{G}$ in the context of $f_{\theta}$. \emph{\textbf{The scoring function prefers $\mathcal{G}^*$ that leads to the similar prediction as the original graph $\mathcal{G}$.}} For example, in SubgraphX~\cite{subgraphx}, the scoring function is designed as the Shapley values. The constraint $|\mathcal{G}'| \leq \Omega$ controls the size of the explanation results, to promote the sparsity of $\mathcal{G}^*$ and avoid trivial solutions such as $\mathcal{G}^*=\mathcal{G}$. 
Another relevant design of explanation algorithms such as GNNExplainer~\cite{gnnExplainer2019} and PGExplainer~\cite{luo2020parameterized} solves the problem:
\begin{equation}
    \mathcal{G}^* = \argmax_{\mathcal{G}'}\, \text{Score}(\mathcal{G}, \mathcal{G}', f_{\theta}) - \lambda \cdot R(\mathcal{G}'), \\
\end{equation}
where the previous explicit constraint is replaced by a regularization term $\lambda \cdot R(\mathcal{G}')$ controlling the complexity (e.g., size) of $\mathcal{G}'$, and $\lambda$ is the hyperparameter. In both GNNExplainer and PGExplainer, the scoring function measures the mutual information between predicting $\mathcal{G}$ and $\mathcal{G}'$.

The proposed detection framework is shown in Figure~\ref{fig:pipeline}.
A graph model $f_{\theta}$ is trained over $\mathcal{D}_{train}$, which serves as the detector. We find that $f_{\theta}$ is prone to depend solely on trigger patterns given backdoor samples, since backdoor triggers are created manually and tend to be strong features learned by neural networks~\cite{composite2020}. On the contrary, clean samples usually depend on more natural and complex features. Therefore, we leverage explanation methods to be used as ``smart" filters, where: 1) the explanation result $\mathcal{G}^*$ for a backdoor sample is very likely to find the full trigger patterns, making $- \bm{y} \cdot log(f_{\theta}(\mathcal{G}))\approx - \bm{y} \cdot log(f_{\theta}(\mathcal{G}^*))$ for $\mathcal{G}\in \mathcal{D}^{poi}_{train}$; 2) the key features for clean samples are not well covered in $\mathcal{G}^*$, so that $- \bm{y} \cdot log(f_{\theta}(\mathcal{G})) < - \bm{y} \cdot log(f_{\theta}(\mathcal{G}^*))$ for $\mathcal{G}\in \mathcal{D}^{clean}_{train}$. 
Such a difference could be leveraged to distinguish between backdoor and clean instances.

\vspace{4pt}
\noindent\textbf{How to Find a Desirable $\mathcal{G}^*$?} 
The motivation of choosing explanation comes from the commonly adopted guidelines of backdoor design~\cite{backdoor_survey,gao2020backdoor}. That is, attackers want the trigger patterns to be stealthy, which means that the size of the trigger pattern should be small. Thus, we could set $\Omega$ as a small value or make $\lambda$ large, so that: 1) backdoor triggers are successfully extracted, and 2) explainable evidence in clean samples is not fully extracted. Then, if we mask the training samples with explanation results, the loss between backdoor samples and clean samples would be distinguished. 
The above motivation is also supported by the following empirical findings: Backdoor samples have a sparse attribution distribution, where the trigger subgraph found by $\mathcal{G}^*$ almost has the same prediction score as the entire graph; in contrast, clean samples have a dispersed attribution distribution, where features not contained in the subgraph $\mathcal{G}^*$ also have non-trivial contributions to the model prediction. 

\vspace{4pt}
\noindent\textbf{How to Find a Desirable $f_{\theta}$?}
Our backdoor detector is based on $f_{\theta}$, which is trained over the dataset $\mathcal{D}_{train}$. A desirable $f_{\theta}$ is expected to well fit the backdoor samples, but mediocrely fit the clean samples. This is because such a model $f_{\theta}$ will amplify the loss gap between backdoor samples and clean samples, making the backdoor samples more isolated in the latent space. To achieve this, we propose the following loss function:
\begin{equation}
    L_{trap} = \mathbb{E}_{(\mathcal{G}, y) \sim \mathcal{D}_{train}}(\ell (f_{\theta}, \mathcal{G}, y) - \gamma)^2,
    \label{eqn:loss}
\end{equation} 
where $\gamma$ is the loss threshold and $\ell(\cdot)$ is the loss function on a single example. The $L_{trap}$ loss function is a quadratic version of the original loss with the axis of symmetry at $\gamma$. When a neural network $f_\theta$ is trained with $L_{trap}$, the loss on most samples will be trapped around $\gamma$. However, because the learning task on backdoor samples is relatively easier, the loss on backdoor samples will not be trapped and can still achieve a low value. In this way, the backdoor samples show distinguished loss values over the clean samples.

\subsection{One-hop Neighbor Expansion}
We further present a simple but effective step, i.e., slightly expanding the extracted subgraph, to increase the likelihood that the trigger is covered by the explanatory subgraph. The neighbor nodes, which are topologically close to the extracted subgraph $\mathcal{G}^*$, are the best candidates to be selected. Formally, for an extracted subgraph $\mathcal{G}^* = (\mathcal{V}^*, \mathcal{E}^*, \mathcal{X}^*)$, it is expanded to a graph $\tilde{\mathcal{G}}=(\tilde{\mathcal{V}}, \tilde{\mathcal{E}}, \tilde{\mathcal{X}})$ with
\begin{equation}
\begin{split}
    &\tilde{\mathcal{V}} = \{v| v \in \mathcal{N}({\mathcal{G}^*}) \cup \mathcal{V}^*\}, \\
    &\tilde{\mathcal{E}} = \{(i,j) | i \in \mathcal{V}^* \vee j \in \mathcal{V}^*, \forall (i, j) \in \mathcal{E}\}, \\
    &\tilde{\mathcal{X}} = \{x| x \in \mathcal{X}(\mathcal{N}({\mathcal{G}^*})) \cup \mathcal{X}(\mathcal{G}^*)\},
    \label{eqn:expansion}
\end{split}
\end{equation}
where $\mathcal{E}$ denotes the edge set of the entire graph.
\subsection{Loss Computation}
Following the above steps, we have obtained a dataset of subgraphs $\hat{\mathcal{D}} = \{(\tilde{\mathcal{G}}_1, y_1), (\tilde{\mathcal{G}}_2, y_2),...,(\tilde{\mathcal{G}}_n, y_n)\}$. Then, we compute the loss value on each sample $(\tilde{\mathcal{G}}_i, y_i)$.
\begin{equation}
    \ell (f_{\theta}, \tilde{\mathcal{G}}_i, y_i) = - \bm{y}_i \cdot log(f_{\theta}(\tilde{\mathcal{G}}_i)),
    \label{eqn:calculate_loss}
\end{equation}
where $\bm{y}_i \in \{0,1\}^{C}$ denotes the one-hot vector with the $y_i$-th component as one and the remaining components as zero.
To identify the backdoor samples, we simply compare the loss on the samples to a predefined threshold $\tau$. The choices for $\tau$ will be shown in the experiment part. The backdoor samples are detected as:
\begin{equation}
    \mathcal{D}_{backdoor} = \{\tilde{\mathcal{G}}_i |  \ell (f_{\theta}, \tilde{\mathcal{G}}_i, y_i) \leq \tau, \forall i \in [1,n] \}.
    \label{eqn:select}
\end{equation}
We formalize the whole process in Algorithm~\ref{alg:detection}.
\begin{algorithm}[tb]
   \caption{The \(\abbr\) backdoor detection framework.}
   \label{alg:detection}
\begin{algorithmic}
   \State {\bfseries Input:} Dataset $\mathcal{D}_{train}=\{(\mathcal{G}_1, y_1), ...,(\mathcal{G}_n, y_n)\}$; detection threshold $\tau$. 
    \State Train a model $f_{\theta}$ over $\mathcal{D}_{train}$ with loss in Equation~\ref{eqn:loss}.
   \For{$i=1$ {\bfseries to} $n$}
   \State Obtain subgraph $\mathcal{G}^*_i$ for each $(\mathcal{G}_i, y_i) \in \mathcal{D}_{train}$.
   \State Expand subgraph $\mathcal{G}_i^*$ to $\tilde{\mathcal{G}}_i$ following Equation~\ref{eqn:expansion}.
   \State Compute the loss Value on $\tilde{\mathcal{G}}_i$ following Equation~\ref{eqn:calculate_loss}.
   \EndFor
   \State Detect backdoor samples following Equation~\ref{eqn:select}.
\end{algorithmic}
\end{algorithm}

\section{Experiments}
In this section, we conduct multiple experiments to 
answer the following research questions. \textbf{RQ1:} How effective is the proposed detection method compared to baseline methods? \textbf{RQ2:} What is the impact of hyperparameters on the effectiveness of the proposed detection method? \textbf{RQ3:} Why is the proposed method effective in detecting backdoor samples? \textbf{RQ4}: What is the computational efficiency of the proposed method?

\subsection{Experimental Settings}
\noindent\textbf{Graph Datasets.}
We evaluate our approach on three real-world graph datasets, where the basic statistics are given in Table~\ref{tab:graph_info}.
\begin{itemize}[leftmargin=*]
    \item \textbf{MUTAG~\cite{mutag}} is a collection of nitroaromatic compounds and the goal is to predict whether they would have a mutagenic effect on Salmonella typhimurium. 
    \item \textbf{AIDS~\cite{aids}}  contains 2000 chemical compounds. The goal is to predict whether the molecules have activity against HIV or not.
    \item \textbf{ENZYMES~\cite{tudataset}} is a dataset containing six different protein tertiary structures. The task is to predict the class of each protein structure. 
\end{itemize}
\begin{table}[!t]
    \caption{Basic statistics of the three datasets.}
    \vspace{-1mm}
    \small
    \centering
    \begin{tabular}{l|cccc}
    \toprule
        Dataset & Graphs & Classes & Avg. Nodes & Avg. Edges \\
        \midrule
        MUTAG & 188 & 2 & 17.93 & 19.79 \\
        \midrule
        AIDS & 2000 & 2 & 15.69 & 16.20 \\
        \midrule
        ENZYMES & 600 & 6 & 32.63 & 62.14 \\
        \bottomrule
    \end{tabular}
    \label{tab:graph_info}
\end{table}

\begin{table*}[!t]
    \caption{The detection accuracy (Accuracy) and area under curve (AUC) of backdoor detection methods against different attacks.}
    \label{tab:main_results}
    \vspace{-2pt}
    \centering
    \resizebox{0.9\textwidth}{!}{%
        \begin{tabular}{c|c|cc|cc|cc|cc}
            \toprule
            \multirow{2}{*}{Dataset} & \multirow{2}{*}{Attack Method} & \multicolumn{2}{c}{AC} & \multicolumn{2}{c}{ABL} &
                 \multicolumn{2}{c}{ExD} & \multicolumn{2}{c}{Ours}\\ 
                 \cmidrule{3 - 10}
             & &  Accuracy  & AUC & Accuracy & AUC  & Accuracy & AUC & Accuracy & AUC  \\
            \midrule
            \multirow{3}{*}{MUTAG} & BadGraph 	& $0.91_{\pm 0.02}$ & - & $0.89_{\pm 0.02}$ & $0.84_{\pm 0.10}$ & $0.44_{\pm 0.18}$ & $0.61_{\pm 0.06}$ &  $\textbf{0.98}_{\pm \textbf{0.02}}$ &  $\textbf{0.94}_{\pm \textbf{0.08}}$\\
            & ExA & $0.91_{\pm 0.01}$ & - & $0.91_{\pm 0.02}$ &  $0.87_{\pm 0.02}$ & $0.60_{\pm 0.24}$ & $0.65_{\pm 0.09}$ & $ \textbf{0.92}_{\pm \textbf{0.04}}$ & $\textbf{0.90}_{\pm \textbf{0.07}}$ \\
            & GTA & $0.90_{\pm 0.02}$  & - & $0.88_{\pm 0.01}$ & $0.76_{\pm 0.02}$ & $0.64_{\pm 0.19}$  & $0.66_{\pm 0.09}$ & $ \textbf{0.97}_{\pm \textbf{0.02}}$ &  $\textbf{0.88}_{\pm \textbf{0.04}}$ \\
            \midrule
            \multirow{3}{*}{AIDS} & BadGraph & $0.91_{\pm 0.03}$ & -  & $0.89_{\pm 0.02}$ &  $\textbf{0.83}_{\pm \textbf{0.13}}$ & $0.88_{\pm 0.04}$ & $0.62_{\pm 0.06}$ & $\textbf{0.93}_{\pm \textbf{0.04}}$ & $0.78_{\pm 0.13}$				\\
            & ExA & $0.92_{\pm 0.03}$ & - & $0.89_{\pm 0.02}$ &  $0.84_{\pm 0.11}$ & $0.91_{\pm 0.03}$ & $0.63_{\pm 0.05}$ & $ \textbf{0.93}_{\pm \textbf{0.01}}$ & $\textbf{0.84}_{\pm \textbf{0.03}}$ \\
            & GTA & $0.88_{\pm 0.03}$ & - & $0.89_{\pm 0.01}$  &  $\textbf{0.84}_{\pm \textbf{0.11}}$ & $0.87_{\pm 0.03}$ & $0.60_{\pm 0.06}$ &  $ \textbf{0.96}_{\pm \textbf{0.01}}$ & $0.74_{\pm 0.03}$\\
            \midrule            
            \multirow{3}{*}{ENZYMES} & BadGraph  & $0.93_{\pm 0.04}$ & - & $0.89_{\pm 0.05}$ & $0.83_{\pm 0.10}$ & $0.92_{\pm 0.03}$ & $0.63_{\pm 0.04}$ &  $\textbf{0.96}_{\pm \textbf{0.02}}$ &  $\textbf{0.96}_{\pm \textbf{0.01}}$ \\
            & ExA & $0.91_{\pm 0.01}$ & - & $0.89_{\pm 0.01}$ &  $0.84_{\pm 0.03}$ & $0.91_{\pm 0.02}$ & $0.58_{\pm 0.02}$ &  $\textbf{0.94}_{\pm \textbf{0.03}}$ & $\textbf{0.87}_{\pm \textbf{0.09}}$ \\
            & GTA & $0.90_{\pm 0.01}$ & - & $0.88_{\pm 0.02}$ & $0.67_{\pm 0.09}$ & $0.76_{\pm 0.13}$ & $0.54_{\pm 0.04}$ &  $\textbf{0.93}_{\pm \textbf{0.04}}$ &  $\textbf{0.88}_{\pm \textbf{0.05}}$\\
            \midrule
        \end{tabular}}
        \begin{scriptsize}
\begin{tablenotes}[leftmargin=*]
    \item \hspace{4mm} - AC uses binary clustering to detect backdoor samples, which does not incorporate any threshold value for computing AUC.
\end{tablenotes}
\end{scriptsize}
\end{table*}

\noindent\textbf{Attack Methods.}
We consider three state-of-the-art backdoor attack methods: BadGraph~\cite{zaixi2021}, GTA~\cite{xi2020}, and Explainability-based Attack~\cite{xu2021}. The configurations for these attack methods are introduced as follows:
\begin{itemize}[leftmargin=*]
    \item \textbf{BadGraph~\cite{zaixi2021}.} This attack involves injecting a subgraph trigger into a graph. For all experiments, the subgraph trigger is generated by the Erdős-Rényi (ER) model~\cite{gilbert1959random}, and the poisoned nodes are randomly selected from the node set of the graph.
    \item \textbf{GTA~\cite{xi2020}.} 
    We train a topology generator and a feature generator for five epochs respectively, which can be utilized to generate a series of candidate backdoor triggers. Then a bi-level optimization problem is solved iteratively to poison the victim graphs with the backdoor triggers. For all the experiments, we set the number of epochs as 20 for the bi-level optimization. 
    \item \textbf{Explainability-based Attack (ExA)~\cite{xu2021}.} 
    We first adopt GraphExplainer~\cite{gnnExplainer2019} to generate a node importance matrix for each victim graph with a pre-trained model. Then the nodes with top-$k$ highest importance values are replaced by the subgraph trigger, where $k$ denotes the number of poisoned nodes.
\end{itemize}

\noindent\textbf{Baseline Defense Methods.}
We select three backdoor defense methods as our defense baselines.
First, we consider Explainability-based Defense (ExD)~\cite{gnnDefense2022}, an existing defense method designed especially for GNN backdoor attacks. Furthermore, we adopt two detection methods developed for image data to the graph domain, including Activation Clustering (AC)~\cite{activation_cluster} and Anti-backdoor Learning (ABL)~\cite{abl}. The details of the defense baselines are given as follows:
\begin{itemize}[leftmargin=*]
    \item \textbf{ExD:} Our implementation follows the algorithms described in~\cite{gnnDefense2022}. Specifically, they adopt explanation methods to generate a subgraph mask and thereby calculate an Explainability Score (ES) to distinguish the backdoor samples and clean samples. We follow the settings in the original paper, adopting GNNExplainer as the explanation method and using an additional validation set $\mathcal{D}_{val} \subset \mathcal{D}_{test}$ to calculate the decision boundary by default.

    \item \textbf{Activation Clustering:} Our implementation follows the detection algorithms described in~\cite{activation_cluster}. Specifically, we select a fully-connected layer to compute the activation values and use an ICA algorithm to perform dimensionality reduction on the activation values. Then we adopt a K-NN algorithm with k = 2 to cluster the activation values. The cluster whose size is relatively smaller will be considered as the backdoor cluster, within which all the samples are detected as backdoor samples. Please note that, according to the original paper, if the size of the smaller cluster is greater than p\%, then all samples will be considered as clean. In all experiments, p is set as 15 by default.
    \item \textbf{ABL:} Our implementation follows the open-source code\footnote{https://github.com/bboylyg/ABL}. Specifically, we train the GNNs with the proposed LGA loss for 20 epochs, which is observed to be able to achieve the optimal isolation precision rate in the Cifar10 dataset. Then we classify the samples with the top 10\% lowest values as the backdoor samples, and the others as the clean samples. Please note that the LGA loss function is formulated as follows:
    \begin{equation}
        LGA = (loss-\gamma) \cdot loss,
    \end{equation}
    where $\gamma$ is set as 0.5 by default and $loss$ is the cross-entropy loss.
\end{itemize}

\noindent\textbf{Evaluation Metrics.}
To evaluate backdoor detection performance, we adopt two commonly used metrics~\cite{Kolouri2020,AEVA2020,dong2021}: 1) Detection Accuracy (Accuracy), which measures the proportion of correctly identified samples; 2) The Area under Receiver Operating Curve (AUC), which measures the trade-off between the detection error rate for clean samples and the detection success rate for backdoor samples.

\subsection{RQ1: Backdoor Detection Performance}
\noindent\textbf{Implementation Details.}
For all attack experiments, we use GIN~\cite{gin2018xu} as the GNN network architecture with a learning rate of 0.01 and training epochs of 100. We use the cross-entropy function as the loss function $\ell$. The default settings for the attack methods follow~\cite{zaixi2021}: trigger density $d = 0.8$ and trigger size $t = 0.2$, where trigger density is used to determine the probability of choosing each edge for the subgraph trigger and trigger size is used to determine the number of nodes in the subgraph trigger. Given a dataset $\mathcal{D}_{train}$ with an average node number $\mathcal{N}_{avg}$, the number of nodes in the subgraph trigger is equal to $\mathcal{N}_{avg} \times \text{trigger size}$. Furthermore, we select the poisoning ratio as $\eta = 10\%$ following~\cite{abl}. For the three evaluated datasets, the loss values on subgraph samples are nearly close to zero, while those on the clean samples are significantly larger, so we choose a small threshold $\tau=1e{-5}$ for the three datasets. The size constraint on the explanatory subgraph is $\Omega = 4$ for MUTAG and AIDS dataset, and $\Omega = 6$ for the ENZYMES dataset. The loss threshold $\gamma$ in Equation~\ref{eqn:loss} is chosen as 0.5. Besides, we adopt subgraphX~\cite{subgraphx} as the explanation method by default. The ablation studies on these hyperparameters will be conducted in later subsections.

\vspace{4pt}
\noindent\textbf{Comparing with Baselines.}
To evaluate the effectiveness of \( \abbr \), we compare it with the three baseline methods. Table~\ref{tab:main_results} presents our main results. As shown, \(\abbr\) achieves a satisfactory performance for all experimental instances. In contrast, despite that other baseline methods sometimes achieve a comparatively higher AUC value than \(\abbr\), the overall performance still implies that they are less effective than \(\abbr\).  Compared to AC and ABL, the advantage of \(\abbr\) comes from the use of topological information in graph data. Specifically, our explanation-based topological feature filter effectively distinguishes between backdoor samples and clean samples. Compared to ExD, the advantage comes from our exploration for the detector function $f_{\theta}$, which effectively enlarges the loss gap with the proposed trap training loss function.

\subsection{RQ2: Ablation Studies}
\noindent\textbf{Impact of Explanation Methods.}
\begin{table}[!t]
    \caption{The detection accuracy (Accuracy) and area under curve (AUC) of \(\abbr\) with different explanation methods.}
    \label{tab:explanation}
    \vspace{-2pt}
    \centering
    \resizebox{0.48\textwidth}{!}{%
        \begin{tabular}{c|c|cc|cc|cc}
            \toprule
            \multirow{2}{*}{Dataset} & \multirow{2}{*}{Attack Method} & \multicolumn{2}{c}{SubgraphX} & \multicolumn{2}{c}{GNNExplainer} & \multicolumn{2}{c}{PGExplainer}\\ 
                 \cmidrule{3 - 8}
             & &  Accuracy  & AUC  & Accuracy & AUC & Accuracy  & AUC \\
            \midrule
            \multirow{3}{*}{MUTAG} & BadGraph & $0.98_{\pm 0.02}$ &  $0.94_{\pm 0.08}$ & $0.97_{\pm 0.03}$ & $0.93_{\pm 0.06}$ & $0.97_{\pm 0.02}$ & $0.93_{\pm 0.03}$\\
            & ExA & $ 0.92_{\pm 0.04}$ & $0.90_{\pm 0.07}$ & $0.96_{\pm 0.01}$ & $0.80_{\pm 0.12}$ & $0.96_{\pm 0.02}$ & $0.81_{\pm 0.10}$ \\
            & GTA  & $ 0.97_{\pm 0.02}$ &  $0.88_{\pm 0.04}$ & $0.94_{\pm 0.03}$ & $0.85_{\pm 0.06}$ & $0.92_{\pm 0.03}$ & $0.92_{\pm 0.01}$ \\
            \midrule
            \multirow{3}{*}{AIDS} & BadGraph & $0.93_{\pm 0.04}$ & $0.78_{\pm 0.13}$ & $0.91_{\pm 0.07}$ & $0.76_{\pm 0.17}$ & $0.93_{\pm 0.01}$ & $0.86_{\pm 0.10}$			\\
            & ExA  & $ 0.93_{\pm 0.01}$ & $0.87_{\pm 0.02}$ & $0.92_{\pm 0.02}$ & $0.74_{\pm 0.02}$ & $0.93_{\pm 0.02}$ & $0.84_{\pm 0.04}$  \\
            & GTA  &  $ 0.96_{\pm 0.01}$ & $0.74_{\pm 0.03}$ & $0.93_{\pm 0.01}$ & $0.73_{\pm 0.05}$ & $0.94_{\pm 0.01}$ & $0.77_{\pm 0.05}$\\
            \midrule            
            \multirow{3}{*}{ENZYMES} & BadGraph  &  $0.96_{\pm 0.02}$ &  $0.96_{\pm 0.01}$ & $0.95_{\pm 0.04}$ & $0.92_{\pm 0.07}$ & $0.95_{\pm 0.05}$ & $0.81_{\pm 0.18}$\\
            & ExA &  ${0.94}_{\pm {0.03}}$ & ${0.87}_{\pm {0.09}}$ & $0.94_{\pm 0.01}$ & $0.80_{\pm 0.12}$ & $0.96_{\pm 0.01}$ & $0.81_{\pm 0.10}$\\
            & GTA &  ${0.93}_{\pm {0.04}}$ &  ${0.88}_{\pm {0.05}}$ & $0.93_{\pm 0.03}$ & $0.85_{\pm 0.06}$ & $0.92_{\pm 0.02}$ & $0.92_{\pm 0.01}$ \\
            \midrule
            & Avg Ranking $\downarrow$ & \textbf{1.67} & \textbf{1.44} & 2.22 & 2.89 & 2.11 & 1.67 \\
            \bottomrule
        \end{tabular}}
\end{table}
We select GNNExplainer~\cite{gnnExplainer2019}, PGExplainer~\cite{luo2020parameterized}, and SubgraphX~\cite{subgraphx} as the explanation methods. Table~\ref{tab:explanation} presents the performance of \(\abbr\) with different explanation methods. To compare the three explanation methods, we calculate their rankings of Accuracy and AUC for each experimental instance and finally report the average ranking value. As shown, \(\abbr\) achieves the highest average ranking for Accuracy and AUC with SubgraphX~\cite{subgraphx}, indicating that SubgraphX could provide a relatively more stable and effective performance. 


\noindent\textbf{Impact of Trigger Size.}
\begin{figure}[!t]
    \includegraphics[width=0.48\textwidth]{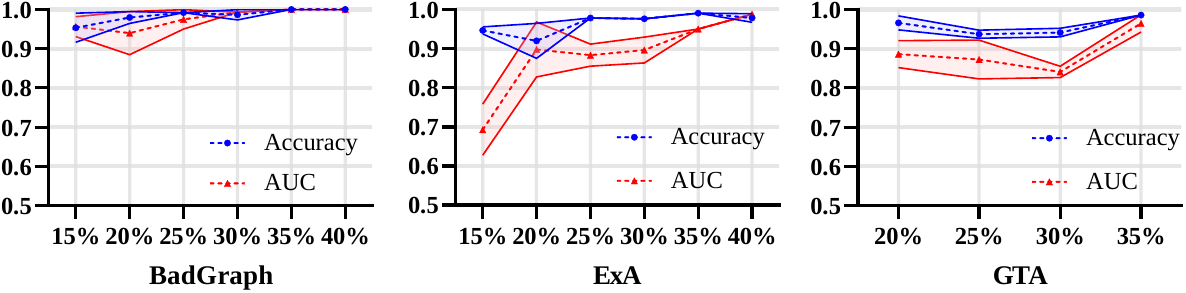}
    \vspace{-10pt}
    \caption{The impact of trigger size.}
    \label{fig:trigger_size}
\end{figure}
Figure~\ref{fig:trigger_size} presents the impact of trigger size under different attack methods in the MUTAG dataset. The X-axis denotes the trigger size, ranging from 15\% to 40\% of the original graph size, and the Y-axis denotes the Accuracy/AUC value. The blue line denotes the accuracy curve and the red line denotes the AUC curve. The standard deviation band is also plotted. We could observe that, for all three attack methods, \(\abbr\) achieves a satisfactory performance when the trigger size is at different levels. Furthermore, the performance of \(\abbr\) tends to be more stable and significant with increasing trigger size. This happens because as the trigger size increases, the subgraph trigger becomes stronger, making the attribution distribution more concentrated on the subgraph trigger.

\begin{figure}[t]
    \centering
    \includegraphics[width=0.48\textwidth]{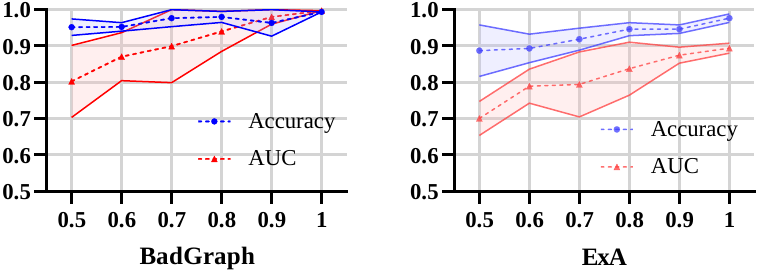}
    \vspace{-10pt}
    \caption{The impact of trigger density.}
    \label{fig:trigger_density}
\end{figure}

\noindent\textbf{Impact of Trigger Density.}
Figure~\ref{fig:trigger_density} presents the impact of trigger density under different attack methods in the MUTAG dataset. The X-axis denotes the trigger density, ranging from 0.5 to 1.0. As GTA~\cite{xi2020} adopts dynamic trigger patterns, we only compare the performance under BadGraph and ExA. As demonstrated, as the trigger density increases, the performance of \(\abbr\) under the two attack methods becomes stronger and more stable. The reasons for this are consistent with the findings of the trigger size impact analysis: the trigger pattern feature becomes stronger.


\noindent\textbf{Impact of Injection Ratio $\eta$.}
\begin{figure}[!t]
    \centering
    \includegraphics[width=0.48\textwidth]{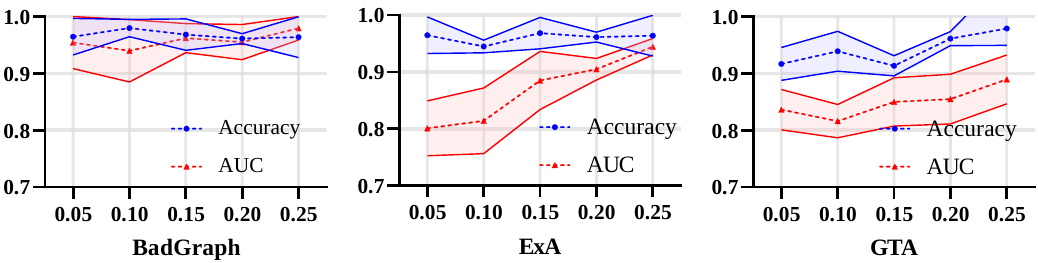}
    \vspace{-10pt}
    \caption{The impact of injection ratio.}
    \label{fig:poisoning_rate}
\end{figure}
Since the number of backdoor samples in the training data is unpredictable, we evaluate the effectiveness of \(\abbr\) under different injection ratios in Figure~\ref{fig:poisoning_rate}. The X-axis denotes the poisoning ratio, and the Y-axis denotes the Accuracy/AUC value. It is shown that our proposed method effectively detects backdoor samples under different poisoning ratios.

\begin{figure}
    \centering
    \includegraphics[width=0.47\textwidth]{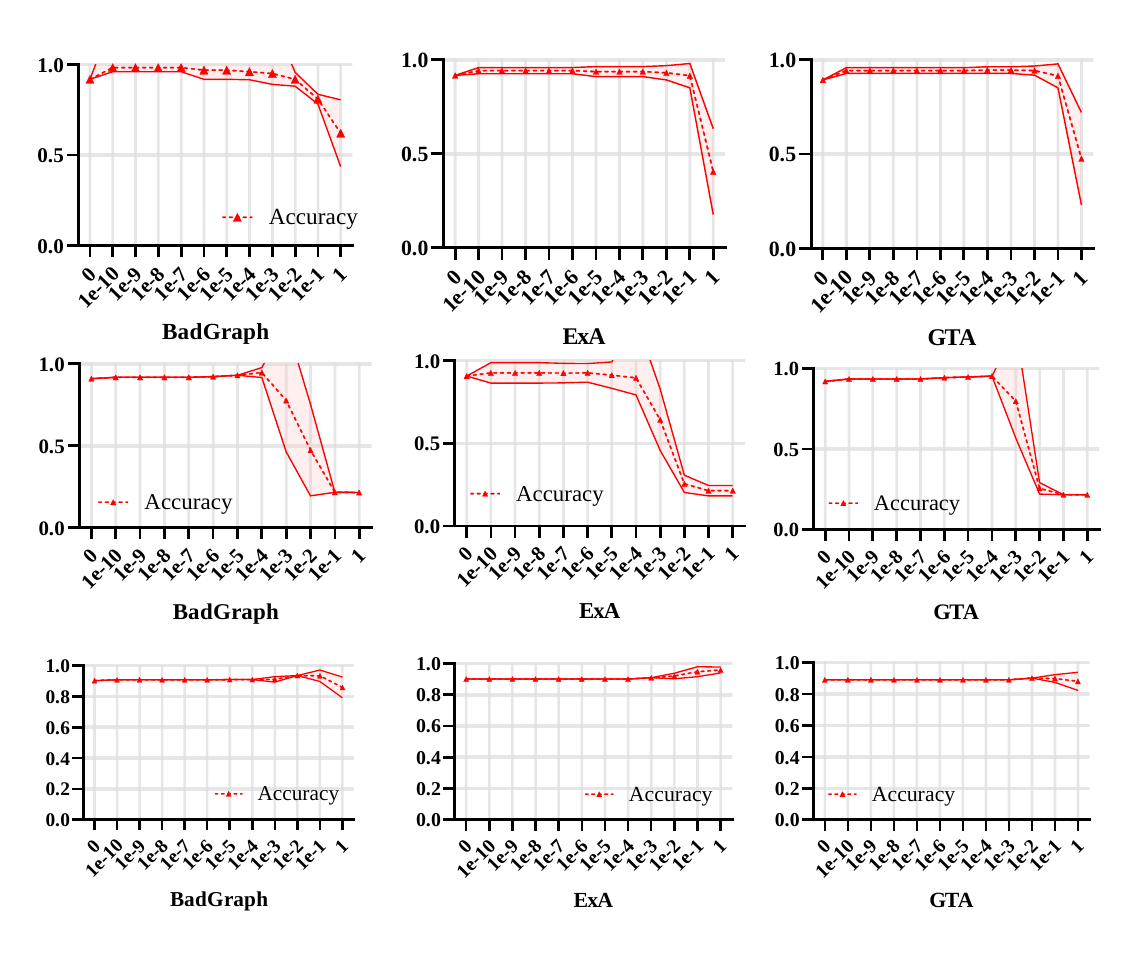}
    \vspace{-3pt}
    \caption{Accuracy with different threshold values $\tau$ over MUTAG dataset (first row), AIDS dataset (second row), and ENZYMES dataset (third row).}
    \label{fig:accuracy}
\end{figure}

\noindent\textbf{Impact of Detection Threshold $\tau$.}
We evaluate the effectiveness of \(\abbr\) under three different attack methods with different thresholds $\tau$ on the three datasets. Figure~\ref{fig:accuracy} presents the experiment results. The X-axis denotes the threshold value, and the Y-axis denotes the accuracy value. \(\abbr\) works well and stably under a wide range of $\tau$ values below 0.01.

\begin{figure}
    \centering
    \includegraphics[width=0.48\textwidth]{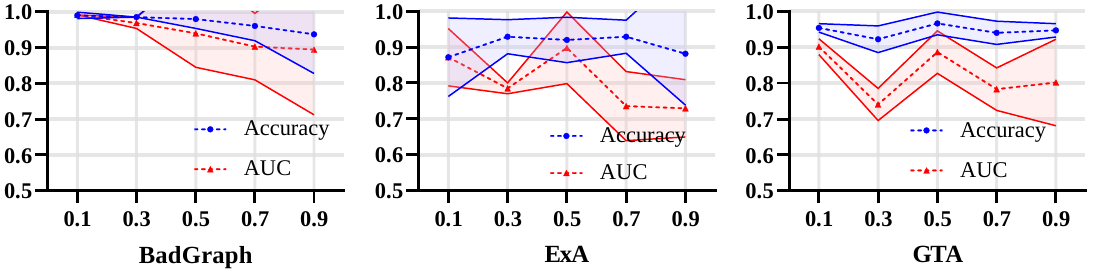}
    \vspace{-10pt}
    \caption{The impact of $\gamma$ value.}
    \label{fig:gamma}
\end{figure}

\noindent\textbf{Impact of Loss Threshold $\gamma$.} Figure~\ref{fig:gamma} presents the performance of \(\abbr\) when training with different $\gamma$ values. As shown in the figure, the Accuracy value and AUC value generally become more unstable with the increase of the $\gamma$ value, and a $\gamma$ value between 0.1 and 0.5 generates a satisfactory result.

\subsection{RQ3: Why is \(\abbr\) Effective?}

\begin{figure}[!t]
    \centering
    \includegraphics[width=0.48\textwidth]{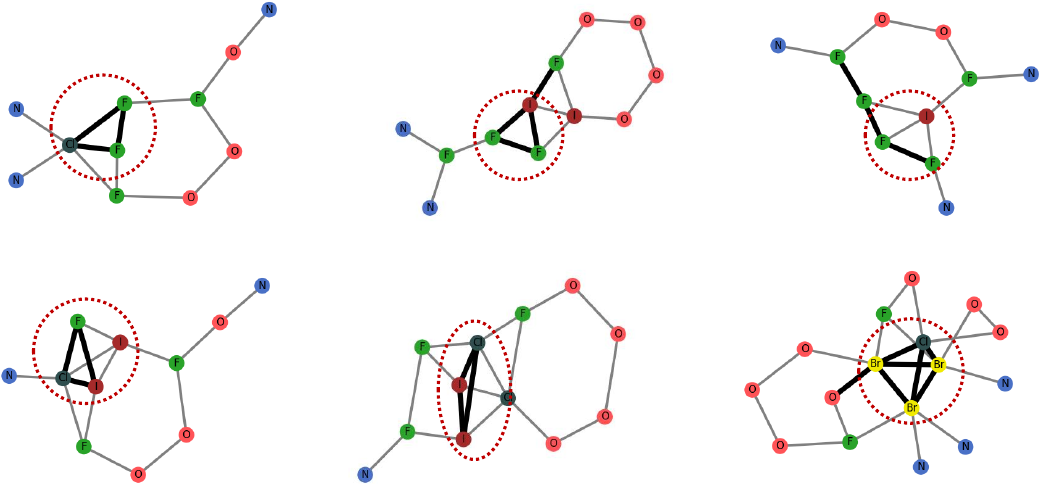}
    \caption{Visualization of identified subgraphs (highlighted by \textbf{bold} lines) by SubgraphX. The true poisoned nodes are in the red circle.}
    \label{fig:viz_subgraphx}
\end{figure}

\noindent\textbf{Explanation Visualizations.}
To better understand how \(\abbr\) helps to detect the backdoor samples, we visualize the results of the explanation method in Figure~\ref{fig:viz_subgraphx}. It presents the identified subgraphs of backdoor samples in the MUTAG dataset. The backdoor samples in the first row are poisoned with a trigger of size three, and those in the second row are poisoned with a trigger of size four. We use red circles to denote poisoned nodes and bold edges to highlight identified subgraphs. As shown, the identified subgraphs align very well with the real poisoned nodes. This is because backdoor models rely almost entirely on the trigger pattern to make predictions when the trigger appears in the input graph.

\begin{figure}[!t]
    \centering
    \includegraphics[width=0.8\columnwidth]{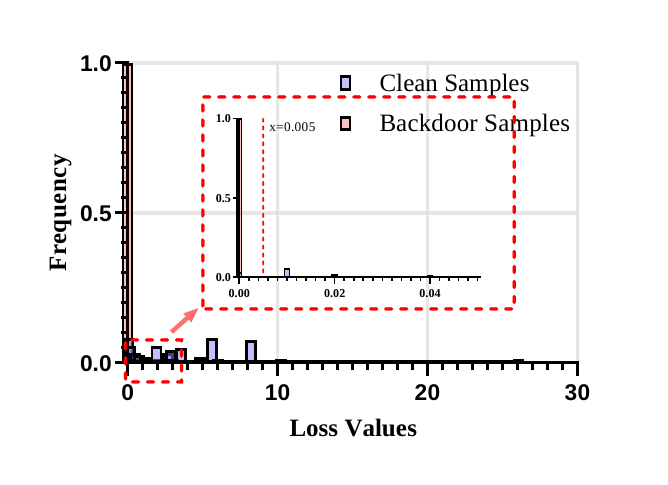}
    \vspace{-5pt}
    \caption{Distribution of loss values on explanatory subgraphs from backdoor samples and clean samples.}
    \label{fig:distribution}
\end{figure}

\noindent\textbf{Loss Distribution.}
Figure~\ref{fig:distribution} presents the distribution of loss values computed from the backdoor subgraphs (blue) and the clean subgraphs (red), respectively. As shown, the loss values of all the backdoor subgraphs are less than 0.005, while the loss values of over 99\% clean subgraphs are higher than 0.005. Therefore, as discussed above, if we set the threshold $\tau$ as a small value such as $1e{-5}$, then the method would effectively separate backdoor and clean samples.

\noindent\textbf{Component Contributions.}
Table~\ref{tab:components} presents the contributions of the \(\abbr\) components. 
The first row represents the scenario in which no explanation methods or subgraph expansion are used, and instead, the training loss is used to identify backdoor samples. The second row corresponds to the scenario in which the explanation method is used, but the subgraph expansion and trap loss are not used. The third row corresponds to the scenario where the explanation method and subgraph expansion are used, but the trap loss is not used.  Finally, the fourth row corresponds to the fully-implemented \(\abbr\). As can be seen, the proposed \(\abbr\) performs better as more components are included, indicating the importance of all three key components.

\begin{table}[!t]
    \caption{The effects of XGBD components on ablation results.}
    \vspace{-2pt}
    \centering
    \resizebox{.48\textwidth}{!}{\begin{tabular}{ccc|cc}
    \toprule
        Explainability Method & Subgraph Expansion & Trap Loss $L_{trap}$ & Accuracy & AUC \\
        \midrule
         - & - & - & $0.91_{\pm 0.04}$ & $0.81_{\pm 0.03}$\\
        \midrule
         \checkmark & -& - & $0.93_{\pm 0.02}$ & $0.92_{\pm 0.03}$ \\
        \midrule
        \checkmark & \checkmark & -
 & $0.95_{\pm {0.05}}$ &  $0.93_{\pm 0.05}$\\
        \midrule
        \checkmark & \checkmark & \checkmark & $\textbf{0.98}_{\pm \textbf{0.02}}$ & $\textbf{0.94}_{\pm \textbf{0.08}}$\\
        \bottomrule
    \end{tabular}}
    \label{tab:components}
\end{table}

\subsection{RQ4: Running Time Efficiency}
In this section, we empirically analyze the running time efficiency of \(\abbr\). Table~\ref{tab:efficiency} compares the average running time of \(\abbr\) with different explanation methods. As shown, when integrating XGBD with an efficient explanation method, such as PGExplainer, the efficiency of XGBD can be guaranteed. For acceleration, the computation on different graph queries could be run in parallel in practice.

\begin{table}[!h]
\small
\caption{Average time (second/sample) consumption.}
\vspace{-3pt}
    \label{tab:efficiency}
    \centering
    \resizebox{0.42\textwidth}{!}{\begin{tabular}{l|ccc}
    \toprule
       Method & MUTAG & AIDS & ENZYMES \\
       \midrule 
        XGBD-PGExplainer & 0.95 & 0.63 & 1.12 \\
        XGBD-GNNExplainer & 6.11 & 5.98 & 6.35\\
        XGBD-SubgraphX & 25.62 & 23.21 & 87.81\\
         \bottomrule
    \end{tabular}}
\end{table}

\section{Related Work}
\subsection{Backdoor Attacks}
Backdoor attacks aim to poison the training dataset with trigger patterns or manipulate the weights and structures of the model directly~\cite{tang2020embarrassingly}. BadNet first proposed backdoor attacks~\cite{gu2017badnets} in the image domain, where they use grid-like patterns as triggers. After training on the poisoned dataset, the backdoor model will continuously predict the target label when the trigger is present in the input image but behaves normally when the input is clean. Following this, many methods are proposed to strengthen attacks by making trigger patterns more stealthy. For example,  ~\cite{clattack} proposes a clean-label attack, which perturbs the clean images without changing their labels.  
Recently, it has been found that backdoor attacks can also be launched in the graph classification task~\cite{zaixi2021,xi2020}.
Simply put, the trigger is a subgraph. When the input graph contains such a subgraph, the prediction would be the target label. When the input graph is clean, the GNN will behave normally. ~\cite{zaixi2021} uses a fixed subgraph to poison the training dataset, where the subgraph is generated randomly and the poisoned nodes of each graph are also generated randomly. ~\cite{xu2021} applies a pre-trained GNN to search for the most vulnerable nodes to be poisoned. ~\cite{xi2020} solves a bi-optimization problem to make the injected subgraph more stealthy.
\subsection{Backdoor Defense}
Various methods have been proposed to defend against backdoor attacks in the image domain. As in~\cite{backdoor_survey}, we categorize existing defense methods into five categories. First, detection-based methods~\cite{strip,2022Xiang} detect whether the backdoor exists in the model by scanning the model prediction or neuron activation values. Second, preprocessing-based methods~\cite{Februs} introduce a data preprocessing module to inactivate backdoors. Third, model reconstruction based defenses~\cite{Fine-pruning,zeng2021adversarial,NEURIPS2021_8cbe9ce2} directly eliminate the effects of backdoors by adjusting weights or structures. Fourth, trigger-synthesis-based defenses~\cite{NeuralCleanse,K_arm_backdoor2021} synthesize the trigger patterns from the victim DNN model and unlearn the patterns learned by the model. Lastly, training sample filtering based defenses~\cite{abl,huang2022backdoor} work by first filtering backdoor samples from the poisoned dataset, then training the network exclusively in the rest of the dataset.

Some of the above methods can be directly adapted to the graph domain~\cite{activation_cluster,abl}. However, limited research has been conducted on the backdoor defenses specially designed for the graph domain. To the best of our knowledge, ~\cite{gnnDefense2022} is the only paper that focuses on removing backdoors in GNNs. In their work, they propose to use an explanation score as the decision criterion for discriminating backdoor samples and clean samples.

\section{Conclusions and Future Work}
This paper tackles the backdoor detection problem in the graph domain, by leveraging explanation methods to identify discriminative topological features. Extensive experiments across three popular graph datasets demonstrate the efficacy of our method across a set of popular tasks and state-of-the-art graph backdoor attacks. Despite the satisfactory performance in detecting backdoor samples, there are still some future works to be done in the research direction. The first one is how to speed up the detection process. Our current detection algorithm sequentially iterates all the training samples, which may take a considerable amount of time when the size of the training dataset becomes significantly large. Therefore, we will consider implementing a parallel version in the future. The second one is how to determine the hyper-parameter: threshold value $\tau$. It would be promising if we determine the value $\tau$ by utilizing some properties of the dataset in an automated way.

\ack
This work is in part supported by NSF (\#IIS-2223768). The views and conclusions contained in this paper are those of the authors and should not be interpreted as representing any funding agencies.

\bibliography{ecai}

\begin{thebibliography}{10}

\bibitem{activation_cluster}
Bryant Chen, Wilka Carvalho, Nathalie Baracaldo, Heiko Ludwig, Benjamin
  Edwards, Taesung Lee, Ian Molloy, and Biplav Srivastava.
\newblock Detecting backdoor attacks on deep neural networks by activation
  clustering, 2018.

\bibitem{chen2023cleanimage}
Kangjie Chen, Xiaoxuan Lou, Guowen Xu, Jiwei Li, and Tianwei Zhang,
  `Clean-image backdoor: Attacking multi-label models with poisoned labels
  only', in {\em The Eleventh International Conference on Learning
  Representations}, (2023).

\bibitem{chou2022backdoor}
Sheng-Yen Chou, Pin-Yu Chen, and Tsung-Yi Ho, `How to backdoor diffusion
  models?', in {\em ICLR 2023 Workshop on Backdoor Attacks and Defenses in
  Machine Learning}, (2022).

\bibitem{mutag}
Asim~Kumar Debnath, Rosa~L. Lopez~de Compadre, Gargi Debnath, Alan~J.
  Shusterman, and Corwin Hansch, `Structure-activity relationship of mutagenic
  aromatic and heteroaromatic nitro compounds. correlation with molecular
  orbital energies and hydrophobicity', {\em Journal of Medicinal Chemistry},
  {\bf 34}(2),  786--797, (1991).

\bibitem{Februs}
Bao~Gia Doan, Ehsan Abbasnejad, and Damith~C. Ranasinghe, `Februus: Input
  purification defense against trojan attacks on deep neural network systems',
  in {\em Annual Computer Security Applications Conference}, ACSAC '20, p.
  897–912, New York, NY, USA, (2020). Association for Computing Machinery.

\bibitem{dong2021}
Yinpeng Dong, Xiao Yang, Zhijie Deng, Tianyu Pang, Zihao Xiao, Hang Su, and Jun
  Zhu.
\newblock Black-box detection of backdoor attacks with limited information and
  data, 2021.

\bibitem{NIPS2017_f5077839}
Alex Fout, Jonathon Byrd, Basir Shariat, and Asa Ben-Hur, `Protein interface
  prediction using graph convolutional networks', in {\em Advances in Neural
  Information Processing Systems}, eds., I.~Guyon, U.~Von Luxburg, S.~Bengio,
  H.~Wallach, R.~Fergus, S.~Vishwanathan, and R.~Garnett, volume~30. Curran
  Associates, Inc., (2017).

\bibitem{gao2019graph}
Hongyang Gao and Shuiwang Ji.
\newblock Graph u-nets, 2019.

\bibitem{gao2020backdoor}
Yansong Gao, Bao~Gia Doan, Zhi Zhang, Siqi Ma, Jiliang Zhang, Anmin Fu, Surya
  Nepal, and Hyoungshick Kim, `Backdoor attacks and countermeasures on deep
  learning: A comprehensive review', {\em arXiv preprint arXiv:2007.10760},
  (2020).

\bibitem{strip}
Yansong Gao, Chang Xu, Derui Wang, Shiping Chen, Damith~C. Ranasinghe, and
  Surya Nepal, `Strip: A defence against trojan attacks on deep neural
  networks', (2019).

\bibitem{gilbert1959random}
Edgar~N Gilbert, `Random graphs', {\em The Annals of Mathematical Statistics},
  {\bf 30}(4),  1141--1144, (1959).

\bibitem{gu2017badnets}
Tianyu Gu, Brendan Dolan-Gavitt, and Siddharth Garg, `Badnets: Identifying
  vulnerabilities in the machine learning model supply chain', {\em arXiv
  preprint arXiv:1708.06733}, (2017).

\bibitem{AEVA2020}
Junfeng Guo, Ang Li, and Cong Liu.
\newblock Aeva: Black-box backdoor detection using adversarial extreme value
  analysis, 2021.

\bibitem{spectre2021}
Jonathan Hayase, Weihao Kong, Raghav Somani, and Sewoong Oh.
\newblock Spectre: Defending against backdoor attacks using robust statistics,
  2021.

\bibitem{hayase2022few}
Jonathan Hayase and Sewoong Oh, `Few-shot backdoor attacks via neural tangent
  kernels', {\em arXiv preprint arXiv:2210.05929}, (2022).

\bibitem{huang2022backdoor}
Kunzhe Huang, Yiming Li, Baoyuan Wu, Zhan Qin, and Kui Ren, `Backdoor defense
  via decoupling the training process', {\em arXiv preprint arXiv:2202.03423},
  (2022).

\bibitem{neuron_inspect}
Xijie Huang, Moustafa Alzantot, and Mani Srivastava.
\newblock Neuroninspect: Detecting backdoors in neural networks via output
  explanations, 2019.

\bibitem{gnnDefense2022}
Bingchen Jiang and Zhao Li.
\newblock Defending against backdoor attack on graph nerual network by
  explainability, 2022.

\bibitem{Kolouri2020}
Soheil Kolouri, Aniruddha Saha, Hamed Pirsiavash, and Heiko Hoffmann.
\newblock Universal litmus patterns: Revealing backdoor attacks in cnns, 2019.

\bibitem{abl}
Yige Li, Xixiang Lyu, Nodens Koren, Lingjuan Lyu, Bo~Li, and Xingjun Ma.
\newblock Anti-backdoor learning: Training clean models on poisoned data, 2021.

\bibitem{backdoor_survey}
Yiming Li, Yong Jiang, Zhifeng Li, and Shu-Tao Xia.
\newblock Backdoor learning: A survey, 2020.

\bibitem{composite2020}
Junyu Lin, Lei Xu, Yingqi Liu, and Xiangyu Zhang, `Composite backdoor attack
  for deep neural network by mixing existing benign features', in {\em
  Proceedings of the 2020 ACM SIGSAC Conference on Computer and Communications
  Security}, CCS '20, p. 113–131, New York, NY, USA, (2020). Association for
  Computing Machinery.

\bibitem{Fine-pruning}
Kang Liu, Brendan Dolan-Gavitt, and Siddharth Garg.
\newblock Fine-pruning: Defending against backdooring attacks on deep neural
  networks, 2018.

\bibitem{luo2020parameterized}
Dongsheng Luo, Wei Cheng, Dongkuan Xu, Wenchao Yu, Bo~Zong, Haifeng Chen, and
  Xiang Zhang, `Parameterized explainer for graph neural network', {\em
  Advances in Neural Information Processing Systems}, {\bf 33}, (2020).

\bibitem{tudataset}
Christopher Morris, Nils~M. Kriege, Franka Bause, Kristian Kersting, Petra
  Mutzel, and Marion Neumann.
\newblock Tudataset: A collection of benchmark datasets for learning with
  graphs, 2020.

\bibitem{aids}
Ryan~A. Rossi and Nesreen~K. Ahmed, `The network data repository with
  interactive graph analytics and visualization', in {\em AAAI}, (2015).

\bibitem{Sanchez2018}
Alvaro Sanchez-Gonzalez, Nicolas Heess, Jost~Tobias Springenberg, Josh Merel,
  Martin Riedmiller, Raia Hadsell, and Peter Battaglia.
\newblock Graph networks as learnable physics engines for inference and
  control, 2018.

\bibitem{clattack}
Ali Shafahi, W.~Ronny Huang, Mahyar Najibi, Octavian Suciu, Christoph Studer,
  Tudor Dumitras, and Tom Goldstein.
\newblock Poison frogs! targeted clean-label poisoning attacks on neural
  networks, 2018.

\bibitem{K_arm_backdoor2021}
Guangyu Shen, Yingqi Liu, Guanhong Tao, Shengwei An, Qiuling Xu, Siyuan Cheng,
  Shiqing Ma, and Xiangyu Zhang.
\newblock Backdoor scanning for deep neural networks through k-arm
  optimization, 2021.

\bibitem{shi2023engage}
Yucheng Shi, Kaixiong Zhou, and Ninghao Liu, `Engage: Explanation guided data
  augmentation for graph representation learning', in {\em ECML-PKDD}, (2023).

\bibitem{tang2020embarrassingly}
Ruixiang Tang, Mengnan Du, Ninghao Liu, Fan Yang, and Xia Hu, `An
  embarrassingly simple approach for trojan attack in deep neural networks', in
  {\em Proceedings of the 26th ACM SIGKDD international conference on knowledge
  discovery \& data mining}, pp. 218--228, (2020).

\bibitem{NeuralCleanse}
Bolun Wang, Yuanshun Yao, Shawn Shan, Huiying Li, Bimal Viswanath, Haitao
  Zheng, and Ben~Y. Zhao, `Neural cleanse: Identifying and mitigating backdoor
  attacks in neural networks', in {\em 2019 IEEE Symposium on Security and
  Privacy (SP)}, pp. 707--723, (2019).

\bibitem{NEURIPS2021_8cbe9ce2}
Dongxian Wu and Yisen Wang, `Adversarial neuron pruning purifies backdoored
  deep models', in {\em Advances in Neural Information Processing Systems},
  eds., M.~Ranzato, A.~Beygelzimer, Y.~Dauphin, P.S. Liang, and J.~Wortman
  Vaughan, volume~34, pp. 16913--16925. Curran Associates, Inc., (2021).

\bibitem{Wu_Lian_Xu_Wu_Chen_2020}
Yongji Wu, Defu Lian, Yiheng Xu, Le~Wu, and Enhong Chen, `Graph convolutional
  networks with markov random field reasoning for social spammer detection',
  {\em Proceedings of the AAAI Conference on Artificial Intelligence}, {\bf
  34}(01),  1054--1061, (Apr. 2020).

\bibitem{wu2020comprehensive}
Zonghan Wu, Shirui Pan, Fengwen Chen, Guodong Long, Chengqi Zhang, and S~Yu
  Philip, `A comprehensive survey on graph neural networks', {\em IEEE
  transactions on neural networks and learning systems}, {\bf 32}(1),  4--24,
  (2020).

\bibitem{xi2020}
Zhaohan Xi, Ren Pang, Shouling Ji, and Ting Wang.
\newblock Graph backdoor, 2020.

\bibitem{2022Xiang}
Zhen Xiang, David~J. Miller, and George Kesidis.
\newblock Post-training detection of backdoor attacks for two-class and
  multi-attack scenarios, 2022.

\bibitem{xu2021}
Jing Xu, Xue Minhui, and Stjepan Picek.
\newblock Explainability-based backdoor attacks against graph neural networks,
  2021.

\bibitem{gin2018xu}
Keyulu Xu, Weihua Hu, Jure Leskovec, and Stefanie Jegelka.
\newblock How powerful are graph neural networks?, 2018.

\bibitem{gnnExplainer2019}
Rex Ying, Dylan Bourgeois, Jiaxuan You, Marinka Zitnik, and Jure Leskovec.
\newblock Gnnexplainer: Generating explanations for graph neural networks,
  2019.

\bibitem{yu2023backdoor}
Yi~Yu, Yufei Wang, Wenhan Yang, Shijian Lu, Yap-peng Tan, and Alex~C Kot,
  `Backdoor attacks against deep image compression via adaptive frequency
  trigger', {\em arXiv preprint arXiv:2302.14677}, (2023).

\bibitem{subgraphx}
Hao Yuan, Haiyang Yu, Jie Wang, Kang Li, and Shuiwang Ji.
\newblock On explainability of graph neural networks via subgraph explorations,
  2021.

\bibitem{zeng2021adversarial}
Yi~Zeng, Si~Chen, Won Park, Zhuoqing Mao, Ming Jin, and Ruoxi Jia, `Adversarial
  unlearning of backdoors via implicit hypergradient', in {\em International
  Conference on Learning Representations}, (2021).

\bibitem{zaixi2021}
Zaixi Zhang, Jinyuan Jia, Binghui Wang, and Neil~Zhenqiang Gong, `Backdoor
  attacks to graph neural networks', in {\em Proceedings of the 26th ACM
  Symposium on Access Control Models and Technologies}, SACMAT '21, p. 15–26,
  New York, NY, USA, (2021). Association for Computing Machinery.

\bibitem{zhang2020deep}
Ziwei Zhang, Peng Cui, and Wenwu Zhu, `Deep learning on graphs: A survey', {\em
  IEEE Transactions on Knowledge and Data Engineering}, (2020).

\bibitem{zhou2020graph}
Jie Zhou, Ganqu Cui, Shengding Hu, Zhengyan Zhang, Cheng Yang, Zhiyuan Liu,
  Lifeng Wang, Changcheng Li, and Maosong Sun, `Graph neural networks: A review
  of methods and applications', {\em AI Open}, {\bf 1},  57--81, (2020).

\end{thebibliography}

\end{document}